\documentclass[12pt,a4paper]{article}
\usepackage{amsmath}
\usepackage{amssymb}
\makeatletter
\renewcommand{\theequation}
{\thesection.\arabic{equation}}
\@addtoreset{equation}{section}
\makeatother

\begin{document}
\begin{flushright}
DPNU-98-43
\end{flushright}

\vspace{30pt}

\begin{center}
{\LARGE Inequivalent Quantization }

{\LARGE in the Skyrme Model\bigskip}

\vspace{30pt}

Hitoshi IKEMORI\footnote{ikemori@biwako.shiga-u.ac.jp} \\[0pt]\textit{Faculty
of Economics, Shiga University,} \textit{Hikone, Shiga 522-8522, Japan}

\medskip

Shinsaku KITAKADO\footnote{kitakado@eken.phys.nagoya-u.ac.jp} \\[0pt]%
\textit{Department of Physics, Nagoya University,} \textit{Nagoya 464-6801,
Japan}

\medskip

Hajime NAKATANI \\[0pt]\textit{Chubu Polytechnic Center, Komaki, Aichi
485-0825, Japan}

\medskip

Hideharu OTSU\footnote{otsu@vega.aichi-u.ac.jp} \\[0pt]\textit{Faculty of
Economics, Aichi University, }

\textit{Toyohashi, Aichi 441-8522, Japan}

\medskip

Toshiro SATO\footnote{tsato@matsusaka-u.ac.jp} \\[0pt]\textit{Faculty of
Political Science and Economics,}\\[0pt]\textit{Matsusaka University,
Matsusaka, Mie 515-0044, Japan}
\end{center}

\vspace*{30pt}

\begin{center}
{\Large Abstract}
\end{center}

Quantum mechanics on manifolds is not unique and in general infinite number of
inequivalent quantizations can be considered. They are specified by the
induced spin and the induced gauge structures on the manifold. The
configuration space of collective mode in the Skyrme model can be identified
with $S^{3}$ and thus the quantization is not unique. This leads to the
different predictions for the physical observables.\newpage

\section{Introduction}

Quantum mechanics on $S^{n}$ is not unique and in general infinite number of
inequivalent quantizations can be considered. They are characterized by the
induced spin \cite{Mackey:1968} and the induced gauge structure
\cite{Landsman:1991zb,Ohnuki:1993cb,McMullan:1995wz}. For example the induced
gauge structure on $S^{1}$ is that of a vortex located at the center of
$S^{1}$ $(r=0)$, for $S^{2}$ the gauge structure is that of a magnetic
monopole also located at the center of $S^{2}$ and furthermore for $S^{3}$
(and for $S^{2n-1}$, $n\geqq3$) the induced gauge field is that of meron or
zero-size instanton (generalized meron or generalized zero-size instanton )
sitting at the center of$\,S^{3}\,(S^{2n-1})$%
\cite{Ikemori:1997fh,Hirayama:1997uf}, while for $S^{4}$ (and for $S^{2n}$,
$n\geqq3$) the gauge field is that of\ instanton (generalized instanton)
\cite{McMullan:1995wz,Fujii:1995wn}. Quantum mechanics on $S^{n}$ has been
formulated in \cite{Ohnuki:1993cb} by extending the canonical commutation
relations that are valid on the flat space, while the authors of
\cite{McMullan:1995wz} have considered $S^{n}$ as a coset space\thinspace
$G/H\,$ and have obtained the effective Lagrangians that correspond to many
inequivalent quantizations. Application of this idea to the physical models
and extension to the field theory
\cite{Tanimura:1994ak,Igarashi:1997tk,Kobayashi:1997rf,Miyazaki:1997yy}\ are
the interesting problems to be pursued.

In this article we consider the SU(2) Skyrme model \cite{Skyrme:1958vn} as a
concrete example, where in the semi-classical approximation the problem
reduces to the quantum mechanics on $S^{3}$. Adkins, Nappi and Witten (ANW)
\cite{Adkins:1983ya} have quantized the system and have shown that the model
describes the static properties of the baryon within 30\% accuracy. We show
that the assignment of the baryonic states in this model is not unique and
that ANW analysis corresponds to the simplest case of inequivalent
quantizations on $S^{3}$. We shall apply the idea of quantum mechanics on
$S^{3}$ to this problem and discuss the effects of induced spin as well as
gauge structure to the physical quantities of this model.

Quantum mechanics on $S^{3}\,$will be reviewed in Sec.2, and Sec.3 is a short
review of SU(2) Skyrme model as developed by ANW. In Sec.4 we shall discuss
the inequivalent quantizations in the Skyrme model. Sec.5 is devoted to
summary and discussions.

\section{Quantum mechanics on $S^{3}$}

In this section we shortly recapitulate the quantum mechanics (QM) on $S^{n}$
\cite{Ohnuki:1993cb}. It is described by the following fundamental
algebra\smallskip%
\begin{align}
\left[  \hat{X}_{\alpha},\hat{X}_{\beta}\right]   & =0,\\
\left[  \hat{X}_{\lambda},\hat{G}_{\alpha\beta}\right]   & =i\left(  \hat
{X}_{\alpha}\delta_{\lambda\beta}-\hat{X}_{\beta}\delta_{\lambda\alpha
}\right)  ,\nonumber\\
\left[  \hat{G}_{\alpha\beta},\hat{G}_{\lambda\mu}\right]   & =i\left(
\delta_{\alpha\lambda}\hat{G}_{\beta\mu}-\delta_{\alpha\mu}\hat{G}%
_{\beta\lambda}+\delta_{\beta\mu}\hat{G}_{\alpha\lambda}-\delta_{\beta\lambda
}\hat{G}_{\alpha\mu}\right)  \;,\nonumber
\end{align}
where $\hat{G}_{\alpha\beta}=-\hat{G}_{\beta\alpha}$ are the generators of
SO($n+1$), and $\hat{X}_{\alpha}$ the coordinates of $S^{n}\;$satisfying
$\sum_{\alpha=1}^{n+1}\hat{X}_{\alpha}^{2}=r^{2}$. It has been noted
\cite{Ohnuki:1993cb} that there exist an infinite number of inequivalent
representations of this algebra, that are characterized by the ``induced
spin'' and ``gauge structure''. This provides a strong contrast to QM on the
flat space, where the irreducible representations of the canonical
algebra\smallskip%
\begin{align}
\left[  \hat{X}_{j},\hat{P}_{k}\right]   & =i\delta_{jk},\\
\left[  \hat{X}_{j},\hat{X}_{k}\right]   & =\left[  \hat{P}_{j},\hat{P}%
_{k}\right]  =0\;,\nonumber
\end{align}
is essentially unique.

In what follows we shall be concerned with the case of $S^{3}$only, the
induced Lagrangian \cite{McMullan:1995wz} of which is given as
\begin{equation}
L_{\mathrm{induced}}=-\text{\textrm{tr}}\left(  Kh^{-1}\dot{h}\right)
-\mathrm{tr}\left(  hKh^{-1}\mathcal{A}_{i}(a)\right)  \dot{a}_{i}%
\;,\label{indued Lagrangian}%
\end{equation}
where $\mathcal{A}_{i}\left(  i=1,2,3\right)  \,$is the induced gauge field
expressed in terms of the $S^{3}$-variable $a_{i\text{ }}$. The variable $h$,
an element of SU(2)$_{\mathrm{V}}$, expresses the induced spin $S_{i}$ as
\begin{equation}
S_{i}=-\mathrm{tr}\left(  T_{i}hKh^{-1}\right)  \;,\label{induced spin}%
\end{equation}
with
\begin{align}
h  & =h_{0}-2T_{i}h_{i}\;\mathbf{,}\\
T_{i}  & =\frac{1}{2i}\left(
\begin{array}
[c]{ll}%
\tau_{i} & 0\\
0 & \tau_{i}%
\end{array}
\right)  \;,\quad\sum_{\alpha=0}^{3}\left(  h_{\alpha}\right)  ^{2}%
=1\;.\nonumber
\end{align}
Here we utilize the coset space representation $G/H$ for the manifold with the
identification $S^{3}=\mathrm{SO}\left(  4\right)  /\mathrm{SO}\left(
3\right)  =\left(  \mathrm{SU}\left(  2\right)  _{\mathrm{L}}\mathrm{\times
SU}\left(  2\right)  _{\mathrm{R}}\right)  /\mathrm{SU}\left(  2\right)
_{\mathrm{V}}.\;K$ is chosen to be $K=kT_{3}$ with a constant $k$. $g\in G$ is
given by
\begin{equation}
g=\left(
\begin{array}
[c]{cc}%
B & 0\\
0 & C
\end{array}
\right)  \;,
\end{equation}
where $B$ and $C\,\ $are $2\times2$ SU(2) matrices.

Using the section $\sigma\left(  a_{i}\right)  $ of $G/H$ expressed as
\begin{equation}
\sigma(a)=gh^{-1}\;,\label{section}%
\end{equation}
\smallskip the induced gauge field ($H$-connection) $\mathcal{A}_{i\text{ }}%
$is given by
\begin{equation}
\mathcal{A}_{i}=\sum_{j=1}^{3}\mathrm{tr}\left(  \sigma^{-1}(a)\partial
_{a_{i}}\sigma(a)T_{j}\right)  T_{j}\;.\label{induced gauge field}%
\end{equation}

The induced gauge field on $S^{3}$ is meron (anti-meron)\cite{Ikemori:1997fh}
or zero-size instanton (anti-instanton)\cite{Hirayama:1997uf} located at the
center of $S^{3}$ in $R^{4}$. The sections corresponding to the meron and
instanton are respectively\smallskip%
\begin{equation}
\sigma=\left(
\begin{array}
[c]{cc}%
1 & 0\\
0 & A
\end{array}
\right)  \;,\quad\sigma=\left(
\begin{array}
[c]{cc}%
A & 0\\
0 & A
\end{array}
\right)
\end{equation}
with
\begin{equation}
A=a_{0}+i\tau_{k}a_{k}\;\mathbf{,\quad}\left(  \sum_{\mu=0}^{3}\left(  a_{\mu
}\right)  ^{2}=1\right)  \;.\label{definition of A}%
\end{equation}
In terms of these the induced Lagrangian is
\begin{equation}
L_{\mathrm{induced}}=-\mathrm{tr}\left(  K\left(  \sigma h\right)  ^{-1}%
\dot{\left(  \sigma h\right)  }\right)  \;.
\end{equation}

\smallskip

\section{Skyrme Model}

In this section we briefly review the SU(2) Skyrme model discussed in
\cite{Adkins:1983ya}. One starts with the Lagrangian
\begin{equation}
\mathcal{L}=\frac{F_{\pi}^{2}}{16}\mathrm{tr}\left(  \partial_{\mu}%
U\partial^{\mu}U^{\dagger}\right)  +\frac{1}{32e^{2}}\mathrm{tr}\left[
\left(  \partial_{\mu}U\right)  U^{\dagger},\left(  \partial_{\nu}U\right)
U^{\dagger}\right]  ^{2}\;,\label{Skyrme Lagrangian}%
\end{equation}
where $U$ is an SU(2) matrix transforming as $U\rightarrow AUB^{-1}$ under the
chiral SU(2)$\times$SU(2) , $F_{\pi}=186$MeV is the pion decay constant. There
are topological soliton solutions in the Lagrangian (\ref{Skyrme Lagrangian})
and we identify the soliton number with the baryon number. To describe the
static soliton , one starts with the ansatz
\begin{equation}
U\left(  \vec{x}\right)  =\exp\left(  iF\left(  r\right)  \vec{\tau}\cdot
\frac{\vec{x}}{r}\right)  \equiv U_{0}\left(  \vec{x}\right)  \;,
\end{equation}
with the boundary conditions
\[
F\left(  r\right)  =\pi\,\;\mathrm{at}\,\;r=0\quad\mathrm{and}\quad F\left(
r\right)  \rightarrow0\;\mathrm{as}\;r\rightarrow\infty\;.
\]
$F\left(  r\right)  \;$is solved numerically by minimizing the static energy.
As a semiclassical approximation to the Skyrme model, ANW consider
\begin{equation}
U\left(  \vec{x},t\right)  =A(t)U_{0}\left(  \vec{x}\right)  A^{\dag
}(t)\;,\label{ANW mode}%
\end{equation}
and quantize the zero mode $A(t)$. Substituting (\ref{ANW mode}) into
(\ref{Skyrme Lagrangian}) and performing the space integral we get
\begin{equation}
L=-M+\lambda\mathrm{tr}\left(  \partial_{0}A\partial_{0}A^{\dag}\right)
\;,\label{ANW Lagrangian}%
\end{equation}
where
\begin{equation}
M=36.5\times\frac{F_{\pi}}{e}%
\end{equation}
is the static energy of the soliton (skyrmion) and
\begin{equation}
\lambda=50.9\times\frac{2\pi}{3}\left(  \frac{1}{e^{3}F_{\pi}}\right)  \;.
\end{equation}
As $A(t)$ is an element of SU(2), (\ref{ANW Lagrangian}) describes a system
defined on $S^{3}\,$(considered in section 2) and thus $A(t)\;$can be
identified with $A\,\ $in (\ref{definition of A}). Isospin transformation of
$A\left(  t\right)  $ is expressed as $A\rightarrow VA$. Space rotation, on
the other hand, can be transferred to the spin transformation of $A(t)$ and is
expressed in terms of $SU\left(  2\right)  $ matrix $R$ as $A\rightarrow AR$.
Thus isospin and spin operators are obtained from the Noether currents
as\smallskip%
\begin{subequations}
\begin{align}
I_{k}  & =\frac{1}{2}i\left(  a_{0}\frac{\partial}{\partial a_{k}}-a_{k}%
\frac{\partial}{\partial a_{0}}-\varepsilon_{klm}a_{l}\frac{\partial}{\partial
a_{m}}\right)  \;,\label{I of ANW}\\
J_{k}  & =\frac{1}{2}i\left(  a_{k}\frac{\partial}{\partial a_{0}}-a_{0}%
\frac{\partial}{\partial a_{k}}-\varepsilon_{klm}a_{l}\frac{\partial}{\partial
a_{m}}\right)  \;,\label{J of ANW}%
\end{align}
where $A=a_{0}+i\tau_{k}a_{k}$ and $\sum_{\mu=0}^{3}a_{\mu}^{2}=1$. Note that
$I^{2}=J^{2}$.

The wave functions for baryons, \textit{i.e.}, the eigenstates of
$(I,I_{3},J,J_{3})$ are
\end{subequations}
\begin{align}
\left|  \mathrm{p}\uparrow\right\rangle  & =\frac{1}{\pi}\left(  a_{1}%
+ia_{2}\right)  ,\quad\left|  \mathrm{p}\downarrow\right\rangle =-\frac{i}%
{\pi}\left(  a_{0}-ia_{3}\right)  \;,\label{ANW baryon wave function}\\
\left|  \mathrm{n}\uparrow\right\rangle  & =\frac{i}{\pi}\left(  a_{0}%
+ia_{3}\right)  ,\quad\left|  \mathrm{n}\downarrow\right\rangle =-\frac{1}%
{\pi}\left(  a_{1}-ia_{2}\right)  \;,\nonumber\\
\left|  \Delta^{++},s_{z}=\frac{3}{2}\right\rangle  & =\frac{\sqrt{2}}{\pi
}\left(  a_{1}+ia_{2}\right)  ^{3}\;,\nonumber\\
\left|  \Delta^{+},s_{z}=\frac{1}{2}\right\rangle  & =-\frac{\sqrt{2}}{\pi
}\left(  a_{1}+ia_{2}\right)  \left\{  1-3\left(  a_{0}^{2}+a_{3}^{2}\right)
\right\}  \;,\cdots.\nonumber
\end{align}

From (\ref{ANW Lagrangian}) the Hamiltonian is
\begin{align}
H  & =M+\frac{1}{2\lambda}J^{2}\\
& =M+\frac{1}{4\lambda}\left(  I^{2}+J^{2}\right)  \;,\nonumber
\end{align}
and the masses of nucleon and delta are
\begin{equation}
M_{\mathrm{N}}=M+\frac{1}{2\lambda}\cdot\frac{3}{4}\;,\quad M_{\Delta}%
=M+\frac{1}{2\lambda}\cdot\frac{15}{4}\;.
\end{equation}
Using the experimental values for $M_{\mathrm{N}}$ and $M_{\Delta}$ one has
\begin{equation}
F_{\pi}=129\mathrm{MeV,}\quad e=5.45\;.
\end{equation}

The magnetic moments of these states are obtained as follows. The isoscalar
and the isovector parts of the magnetic moments are respectively
\begin{subequations}
\begin{align}
\left(  \mu_{I=0}\right)  _{i}  & \equiv\frac{1}{2}\int\varepsilon_{ijk}%
x_{j}B_{k}d^{3}x\;,\label{mu of I=0}\\
\left(  \mu_{I=1}\right)  _{i}  & \equiv\frac{1}{2}\int\varepsilon_{ijk}%
x_{j}V_{k}^{3}d^{3}x\;,
\end{align}
here
\end{subequations}
\begin{equation}
B_{i}=i\frac{\varepsilon_{ijk}}{2\pi^{2}}\frac{\sin^{2}F}{r}F^{\prime}\hat
{x}_{k}\mathrm{tr}\left\{  \left(  \partial_{0}A^{\dag}\right)  A\tau
_{j}\right\} \label{B i}%
\end{equation}
is the space component of the baryon number (topological number) current and
$V_{i}^{3}$ is the space component of the isovector current which satisfies
the relation
\begin{equation}
\int d\Omega x_{k}V_{i}^{a}=\frac{1}{3}i\pi\Lambda\mathrm{tr}\left(  \tau
_{k}\tau_{i}A^{\dag}\tau^{a}A\right)  \;.
\end{equation}
To calculate $\mu_{I=0}$ , one substitutes (\ref{B i}) into (\ref{mu of I=0}),
then carrying out the space integral one is lead to
\begin{align}
\left(  \mu_{I=0}\right)  _{3}  & =i\frac{0.09}{2\pi}\frac{e}{F_{\pi}}%
\lambda\left\langle \mathrm{p}\uparrow\right|  \mathrm{tr}\left(  \partial
_{0}A^{\dag}A\tau_{3}\right)  \left|  \mathrm{p}\uparrow\right\rangle
\label{ANW mu of I=0}\\
& =\frac{0.09}{2\pi}\frac{e}{F_{\pi}}\left\langle \mathrm{p}\uparrow\right|
J_{3}\left|  \mathrm{p}\uparrow\right\rangle \nonumber\\
& =\frac{0.09}{2\pi}\frac{e}{F_{\pi}}\times\frac{1}{2}\nonumber\\
& =3.0\times10^{-4}\;.\nonumber
\end{align}
Here we used the relation
\begin{equation}
\lambda\mathrm{tr}\left(  \partial_{0}A^{\dag}A\tau_{3}\right)  =-iJ_{3}\;.
\end{equation}
Similarly, calculation of $\mu_{I=1}$ is reduced to that of $\left\langle
\mathrm{p}\uparrow\right|  \mathrm{tr}\left(  \tau_{3}A^{\dag}\tau
_{3}A\right)  \left|  \mathrm{p}\uparrow\right\rangle $ and using
\begin{equation}
\left\langle \mathrm{N}^{\prime}\right|  \mathrm{tr}\left(  \tau_{i}A^{\dag
}\tau_{j}A\right)  \left|  \mathrm{N}\right\rangle =-\frac{2}{3}\left\langle
\mathrm{N}^{\prime}\right|  \sigma_{i}\tau_{j}\left|  \mathrm{N}\right\rangle
\;,
\end{equation}
which is valid for the nucleonic states, one finally has
\begin{align}
\left(  \mu_{I=1}\right)  _{3}  & =-50.9\times\frac{\pi}{3}\left(  \frac
{1}{e^{3}F_{\pi}}\right)  \left\langle \mathrm{p}\uparrow\right|
\mathrm{tr}\left(  \tau_{3}A^{\dag}\tau_{3}A\right)  \left|  \mathrm{p}%
\uparrow\right\rangle \label{ANW mu of I=1}\\
& =-50.9\times\frac{\pi}{3}\left(  \frac{1}{e^{3}F_{\pi}}\right)  \left(
-\frac{2}{3}\right)  \left\langle \mathrm{p}\uparrow\right|  \sigma_{3}%
\tau_{3}\left|  \mathrm{p}\uparrow\right\rangle \nonumber\\
& =\frac{2}{9}\pi\frac{50.9}{e^{3}F_{\pi}}\nonumber\\
& =1.7\times10^{-3}\;.\nonumber
\end{align}
From (\ref{ANW mu of I=0}) and (\ref{ANW mu of I=1}) the magnetic moments of
proton and neutron in terms of the Bohr magneton are $\mu_{\mathrm{p}}=1.87$
and $\mu_{\mathrm{n}}=-1.31$ respectively.

ANW have calculated also other physical quantities like $g_{A},$ $g_{\pi
\pi\mathrm{N}}$ and $g_{\pi\mathrm{N}\Delta}$. Their conclusion is that the
model describes the reality within about 30\%.

\section{Inequivalent Quantizations}

As we saw in the previous section the configuration space of the semiclassical
approximation to the Skyrme model is $S^{3}.$ The question of inequivalent
quantizations, discussed in sec.2, arises here. It is clear that the ANW
analysis of the previous section corresponds to the trivial quantization with
the ``induced spin'' $S=0$ and induced gauge field $\mathcal{A}_{\mu}=0$. In
this section we shall discuss the possibility of the other non-trivial
quantizations with non-zero ``induced spin'' and induced gauge field and
examine the physical effects to the results of the Skyrme model. In the
following we shall consider the cases when the induced gauge field
configuration is 1) that of meron and 2) that of zero-size instanton .

\subsection{Meron case}

The section $\sigma(a)$ for the case of meron configuration is
\begin{equation}
\sigma=\left(
\begin{array}
[c]{cc}%
1 & 0\\
0 & A
\end{array}
\right)  =\left(
\begin{array}
[c]{cc}%
1 & 0\\
0 & a_{0}+i\tau_{k}a_{k}%
\end{array}
\right) \label{meron section}%
\end{equation}
As quantum mechanics on $S^{3}$ we introduce the ``induced spin'' and the
induced gauge field by substituting (\ref{meron section}) into (\ref{induced
gauge field}) and (\ref{indued Lagrangian}). Then the resulting effective
Lagrangian is
\begin{align}
L  & =L_{0}+L_{\mathrm{induced}}\label{meron Lagrangian}\\
& =-M+\lambda\mathrm{tr}\left(  \partial_{0}A\partial_{0}A^{\dag}\right)
-\mathrm{tr}\left(  Kh^{-1}\dot{h}\right)  +\mathcal{B}_{\mu}\dot{a}_{\mu
}\;,\nonumber
\end{align}
where
\begin{equation}
\mathcal{B}_{0}=S_{i}a_{i},\quad\mathcal{B}_{i}=-S_{i}a_{0}-\varepsilon
_{ijk}S_{j}a_{k}\;,\label{definition of B}%
\end{equation}
and $S_{i}$ is the ``induced spin'' operator (\ref{induced spin}). As the
dynamical variables we consider $h_{\mu}$ and $a_{\mu}$ . Since the
non-trivial quantization has been already taken into account in (\ref{meron
Lagrangian}), we can carry out the constrained quantization \textit{\`{a} la}
Dirac (see Appendix). We obtain the following commutation relations
\begin{align}
\left[  \hat{a}_{\mu},\hat{a}_{\nu}\right]   &
=0\;,\label{commutation ralations}\\
\left[  \hat{a}_{\mu},\hat{\pi}_{\nu}\right]   & =i\left(  \delta_{\mu\nu
}-\hat{a}_{\mu}\hat{a}_{\nu}\right)  \;,\nonumber\\
\left[  \hat{\pi}_{\mu},\hat{\pi}_{\nu}\right]   & =i\left(  \hat{a}_{\nu}%
\hat{\pi}_{\mu}-\hat{a}_{\mu}\hat{\pi}_{\nu}\right)  \;,\nonumber\\
\left[  \hat{S}_{i},\hat{S}_{j}\right]   & =i\varepsilon_{ijk}\hat{S}%
_{k}\;,\nonumber\\
\left[  \hat{a}_{\mu},\hat{S}_{i}\right]   & =\left[  \hat{\pi}_{\mu},\hat
{S}_{i}\right]  =0\;.\nonumber
\end{align}
$\hat{\pi}_{\mu}$ is canonical conjugate of $\hat{a}_{\mu}$ and in the
$\hat{a}_{\mu}$-diagonal representation is given as
\begin{equation}
\hat{\pi}_{\mu}=-i\left(  \frac{\partial\;}{\partial a_{\mu}}-a_{\mu}a_{\nu
}\frac{\partial\;}{\partial a_{\nu}}\right)  \;.\label{conjugate momentum}%
\end{equation}

Although the expression (\ref{conjugate momentum}) is identical with the one
given in ANW, from
\begin{equation}
\pi_{\mu}=\frac{\partial L}{\partial\dot{a}_{\mu}}=4\lambda\dot{a}_{\mu
}+\mathcal{B}_{\mu}\;,
\end{equation}
we have
\begin{equation}
\dot{a}_{\mu}=\frac{1}{4\lambda}\left(  \pi_{\mu}-\mathcal{B}_{\mu}\right)
\;,\label{a dot (meron)}%
\end{equation}
which depends on the ``induced spin'' and induced gauge field in contrast to
ANW.

From
\[
L_{\mathrm{induced}}=-\mathrm{tr}\left(  K\left(  \sigma h\right)  ^{-1}%
\dot{\left(  \sigma h\right)  }\right)  \;,
\]
isospin and spin transformations are respectively
\begin{subequations}
\begin{align}
& \quad\;\left.  A\rightarrow VA\right.  \;,\\
& \left\{
\begin{array}
[c]{l}%
A\rightarrow AR\\
h\rightarrow\left(
\begin{array}
[c]{cc}%
R^{-1} & 0\\
0 & R^{-1}%
\end{array}
\right)  h
\end{array}
\right.  \;.
\end{align}
Thus using the Noether currents isospin and spin operators are
\end{subequations}
\begin{subequations}
\begin{align}
I_{i}  & =I_{i}^{\mathrm{ANW}}\;,\label{isospin operator I}\\
J_{i}  & =J_{i}^{\mathrm{ANW}}+S_{i}\;,\label{spin operator J}%
\end{align}
where $I_{i}^{\mathrm{ANW}}$ and $J_{i}^{\mathrm{ANW}}$ are the expressions of
ANW given in (\ref{I of ANW}) and (\ref{J of ANW}).

As is seen, spin $J_{i}$ depends on the ``induced spin'' $S_{i}$, thus the
eigen states ($I,J$) are fixed according to the representations of $S$. The
case of ANW corresponds to $S=0$. The only other possibility that can describe
N=$\left(  \frac{1}{2},\frac{1}{2}\right)  $ and $\Delta=(\frac{3}{2},\frac
{3}{2})$ is the case of $S=1$. This case is capable of describing the
following states:
\end{subequations}
\begin{equation}
\left(  I,J\right)  =\left(  \frac{1}{2},\frac{1}{2}\right)  ,\,\left(
\frac{3}{2},\frac{3}{2}\right)  ,\,\left(  \frac{1}{2},\frac{3}{2}\right)
,\,\left(  \frac{3}{2},\frac{1}{2}\right)  ,\,\left(  \frac{3}{2},\frac{5}%
{2}\right)  ,\,\cdots\;.
\end{equation}
In what follows we shall be concerned with the effects of the ``induced spin''
and the induced gauge fields on the Skyrme model, identifying the N and
$\Delta$ with the ($\frac{1}{2},\frac{1}{2})$ and $(\frac{3}{2},\frac{3}{2})$
states in $S=1$.

Baryonic wave functions in our case can be expressed in terms of those of ANW
as
\begin{align}
\left|  \left|  \mathrm{p}\uparrow\right\rangle \right\rangle  & =\frac
{1}{\sqrt{3}}\left(
\begin{array}
[c]{c}%
\sqrt{2}\left|  \mathrm{p}\downarrow\right\rangle \\
-\left|  \mathrm{p}\uparrow\right\rangle \\
0
\end{array}
\right)  ,\quad\left|  \left|  \mathrm{p}\downarrow\right\rangle \right\rangle
=\frac{1}{\sqrt{3}}\left(
\begin{array}
[c]{c}%
0\\
\left|  \mathrm{p}\downarrow\right\rangle \\
-\sqrt{2}\left|  \mathrm{p}\uparrow\right\rangle
\end{array}
\right)  \;,\\
\left|  \left|  \mathrm{n}\uparrow\right\rangle \right\rangle  & =\frac
{1}{\sqrt{3}}\left(
\begin{array}
[c]{c}%
\sqrt{2}\left|  \mathrm{n}\downarrow\right\rangle \\
-\left|  \mathrm{n}\uparrow\right\rangle \\
0
\end{array}
\right)  ,\quad\left|  \left|  \mathrm{n}\downarrow\right\rangle \right\rangle
=\frac{1}{\sqrt{3}}\left(
\begin{array}
[c]{c}%
0\\
\left|  \mathrm{n}\downarrow\right\rangle \\
-\sqrt{2}\left|  \mathrm{n}\uparrow\right\rangle
\end{array}
\right)  \;,\nonumber
\end{align}%
\[
\left|  \left|  \Delta^{++},s_{z}=\frac{3}{2}\right\rangle \right\rangle
=\frac{1}{\sqrt{5}}\left(
\begin{array}
[c]{c}%
-\sqrt{2}\left|  \Delta^{++},s_{z}=\frac{1}{2}\right\rangle \\
\sqrt{3}\left|  \Delta^{++},s_{z}=\frac{3}{2}\right\rangle \\
0
\end{array}
\right)  ,\quad\cdots\;,
\]
where $\left|  \mathrm{p}\uparrow\right\rangle $, $\left|  \mathrm{n}%
\uparrow\right\rangle ,\cdots$ are given in (\ref{ANW baryon wave function}).
The Hamiltonian
\begin{equation}
H=M+\frac{1}{8\lambda}\left(  \pi_{\mu}-\mathcal{B}_{\mu}\right)  ^{2}%
\end{equation}
expressed in terms of $J_{i}=J_{i}^{\mathrm{ANW}}+S_{i}$ and $I_{i}$ is
\begin{equation}
H=M+\frac{1}{4\lambda}\left(  I^{2}+J^{2}-\frac{1}{2}S^{2}\right)
\;,\label{Hamiltonian (meron)}%
\end{equation}
thus the baryon mass does have the effect of the ``induced spin''. From
(\ref{Hamiltonian (meron)}) we have
\begin{align}
H_{\mathrm{N}}  & =M+\frac{1}{2\lambda}\cdot\frac{1}{4}\;,\nonumber\\
H_{\Delta}  & =M+\frac{1}{2\lambda}\cdot\frac{13}{4}\;,
\end{align}
and using the experimental N and $\Delta$ masses we obtain
\begin{align}
F_{\pi}  & =135\,\mathrm{MeV\quad}\left(  \mathrm{ANW}^{\prime}%
\mathrm{s\;value\;}F_{\pi}=129\,\mathrm{MeV}\right)  \;,\\
e  & =5.37\mathrm{\quad}\left(  \mathrm{ANW}^{\prime}\mathrm{s\;value\;}%
e=5.45\right)  \;.\nonumber
\end{align}
Note that the value of $F_{\pi}e^{3}$ is the same as that of ANW.

Next we examine the magnetic moments. As baryon number current and vector
current are the same as in ANW, calculation of ($\mu_{I=0})_{3}\;$and
($\mu_{I=1})_{3}$ reduces to that of the matrix elements $\left\langle
\left\langle \mathrm{p}\uparrow\right|  \right|  \mathrm{tr}\left(
\partial_{0}A^{\dag}A\tau_{3}\right)  \left|  \left|  \mathrm{p}%
\uparrow\right\rangle \right\rangle $, $\left\langle \left\langle
\mathrm{p}\uparrow\right|  \right|  \mathrm{tr}\left(  \tau_{3}A^{\dag}%
\tau_{3}A\right)  \left|  \left|  \mathrm{p}\uparrow\right\rangle
\right\rangle .$ Using (\ref{a dot (meron)}), (\ref{conjugate momentum}) and
(\ref{spin operator J}) we have
\begin{align*}
\mathrm{tr}\left(  \partial_{0}A^{\dag}A\tau_{3}\right)   & =2i\left(
-a_{0}\dot{a}_{3}+a_{3}\dot{a}_{0}-a_{1}\dot{a}_{2}+a_{2}\dot{a}_{1}\right) \\
& =-\frac{i}{\lambda}\left(  J_{3}^{\mathrm{ANW}}+\frac{1}{2}S_{3}\right)  \;,
\end{align*}
consequently
\begin{align}
& \left\langle \left\langle \mathrm{p}\uparrow\right|  \right|  \lambda
\mathrm{tr}\left(  \partial_{0}A^{\dag}A\tau_{3}\right)  \left|  \left|
\mathrm{p}\uparrow\right\rangle \right\rangle \label{mu of I=0 (meron)}\\
& =-i\frac{1}{\sqrt{3}}\left(  \sqrt{2}\left\langle \mathrm{p}\downarrow
\right|  ,\,-\left\langle \mathrm{p}\uparrow\right|  ,\,0\right)  \left(
J_{3}^{\mathrm{ANW}}+\frac{1}{2}S_{3}\right)  \frac{1}{\sqrt{3}}\left(
\begin{array}
[c]{c}%
\sqrt{2}\left|  \mathrm{p}\downarrow\right\rangle \\
-\left|  \mathrm{p}\uparrow\right\rangle \\
0
\end{array}
\right) \nonumber\\
& =-i\frac{1}{3}\left\{  2\left\langle \mathrm{p}\downarrow\right|
J_{3}^{\mathrm{ANW}}\left|  \mathrm{p}\downarrow\right\rangle +\left\langle
\mathrm{p}\uparrow\right|  J_{3}^{\mathrm{ANW}}\left|  \mathrm{p}%
\uparrow\right\rangle +\left\langle \mathrm{p}\downarrow\right|  \left.
\mathrm{p}\downarrow\right\rangle \right\} \nonumber\\
& =-\frac{i}{6}\;.\nonumber
\end{align}
Thus
\begin{align}
\left(  \mu_{I=0}\right)  _{3}  & =i\frac{0.09}{2\pi}\frac{e}{F_{\pi}}%
\lambda\left\langle \left\langle \mathrm{p}\uparrow\right|  \right|
\mathrm{tr}\left(  \partial_{0}A^{\dag}A\tau_{3}\right)  \left|  \left|
\mathrm{p}\uparrow\right\rangle \right\rangle \\
& =\frac{0.09}{2\pi}\frac{e}{F_{\pi}}\times\frac{1}{6}\nonumber\\
& =\frac{0.09}{2\pi}\left(  \frac{e}{F_{\pi}}\right)  ^{\mathrm{ANW}}%
\times0.95\times\frac{1}{6}\nonumber\\
& =\left(  \mu_{I=0}\right)  _{3}^{\mathrm{ANW}}\times0.32\;.\nonumber
\end{align}
Similarly the isovector part is
\begin{align}
\left(  \mu_{I=1}\right)  _{3}  & =-50.9\times\frac{\pi}{3}\left(  \frac
{1}{e^{3}F_{\pi}}\right)  \left\langle \left\langle \mathrm{p}\uparrow\right|
\right|  \mathrm{tr}\left(  \tau_{3}A^{\dag}\tau_{3}A\right)  \left|  \left|
\mathrm{p}\uparrow\right\rangle \right\rangle \label{mu of I=1 (meron)}\\
& =-50.9\times\frac{\pi}{3}\left(  \frac{1}{e^{3}F_{\pi}}\right) \nonumber\\
& \times\left(  -\frac{2}{3}\right)  \frac{1}{\sqrt{3}}\left(  \sqrt
{2}\left\langle \mathrm{p}\downarrow\right|  ,\,-\left\langle \mathrm{p}%
\uparrow\right|  ,\,0\right)  \left(  \sigma_{3}\tau_{3}\right)  \frac
{1}{\sqrt{3}}\left(
\begin{array}
[c]{c}%
\sqrt{2}\left|  \mathrm{p}\downarrow\right\rangle \\
-\left|  \mathrm{p}\uparrow\right\rangle \\
0
\end{array}
\right) \nonumber\\
& =\frac{2}{9}\pi\frac{50.9}{e^{3}F_{\pi}}\frac{1}{3}\left\{  2\left\langle
\mathrm{p}\downarrow\right|  \sigma_{3}\tau_{3}\left|  \mathrm{p}%
\downarrow\right\rangle +\left\langle \mathrm{p}\uparrow\right|  \sigma
_{3}\tau_{3}\left|  \mathrm{p}\uparrow\right\rangle \right\} \nonumber\\
& =\frac{2}{9}\pi\frac{50.9}{e^{3}F_{\pi}}\left(  -\frac{1}{3}\right)
\nonumber\\
& =\left(  \mu_{I=1}\right)  _{3}^{\mathrm{ANW}}\times\left(  -\frac{1}%
{3}\right)  \;.\nonumber
\end{align}
As a result, the values of $\left(  \mu_{I=0}\right)  _{3}$ and $\left(
\mu_{I=1}\right)  _{3}$ differ from those of ANW by the factors 0.32
and$\;-\frac{1}{3}$.

\subsection{Instanton case}

The section $\sigma(a)$ for the zero-size instanton configuration is
\[
\sigma=\left(
\begin{array}
[c]{cc}%
A & 0\\
0 & A
\end{array}
\right)  =\left(
\begin{array}
[c]{cc}%
a_{0}+i\tau_{k}a_{k} & 0\\
0 & a_{0}+i\tau_{k}a_{k}%
\end{array}
\right)  \;.
\]
The effective Lagrangian is given as
\begin{align}
L  & =L_{0}+L_{\mathrm{induced}}\\
& =-M+\lambda\mathrm{tr}\left(  \partial_{0}A\partial_{0}A^{\dag}\right)
-\mathrm{tr}\left(  Kh^{-1}\dot{h}\right)  +\mathcal{\tilde{B}}_{\mu}\dot
{a}_{\mu}\;,\nonumber
\end{align}
where\ $\mathcal{\tilde{B}}_{\mu}=2\mathcal{B}_{\mu}$ .

Commutation relations, isospin and spin operators, baryon wave functions are
the same as in the meron case, the only change being
\begin{equation}
\dot{a}_{\mu}=\frac{1}{4\lambda}\left(  \pi_{\mu}-\mathcal{\tilde{B}}_{\mu
}\right)  =\frac{1}{4\lambda}\left(  \pi_{\mu}-2\mathcal{B}_{\mu}\right)
\label{a dot (instanton)}%
\end{equation}
i.e., the contribution from the ``induced spin'' and the induced gauge field
to $\dot{a}_{\mu}$ is different by a factor 2.

The Hamiltonian expressed in terms of $J_{i}$ and $I_{i}$ is
\begin{align}
H  & =M+\frac{1}{8\lambda}\left(  \pi_{\mu}-\mathcal{\tilde{B}}_{\mu}\right)
^{2}\label{Hamiltonian (instanton)}\\
& =M+\frac{1}{8\lambda}\left(  \pi_{\mu}-2\mathcal{B}_{\mu}\right)
^{2}\nonumber\\
& =M+\frac{1}{4\lambda}\left(  I^{2}+J^{2}\right)  \;,\nonumber
\end{align}
which unlike the case of meron has no contribution from the ``induced spin''.
N and $\Delta$ masses follow from (\ref{Hamiltonian (instanton)})
\begin{align}
H_{\mathrm{N}}  & =M+\frac{1}{2\lambda}\cdot\frac{3}{4}\;,\\
H_{\Delta}  & =M+\frac{1}{2\lambda}\cdot\frac{15}{4}\;,\nonumber
\end{align}
and if we use the experimental values as the input, we obtain
\begin{equation}
F_{\pi}=129\,\mathrm{MeV,\quad}e=5.45\nonumber
\end{equation}
in agreement with ANW.

From (\ref{a dot (instanton)}) we have%
\begin{align}
\mathrm{tr}\left(  \partial_{0}A^{\dag}A\tau_{3}\right)    & =2i\left(
-a_{0}\dot{a}_{3}+a_{3}\dot{a}_{0}-a_{1}\dot{a}_{2}+a_{2}\dot{a}_{1}\right)
\\
& =-\frac{i}{\lambda}\left(  J_{3}^{\mathrm{ANW}}+S_{3}\right)  \nonumber\\
& =-\frac{i}{\lambda}J_{3}\;,\nonumber
\end{align}
and the isoscalar part of magnetic moments is
\begin{align}
\left(  \mu_{I=0}\right)  _{3}  & =i\frac{0.09}{2\pi}\frac{e}{F_{\pi}}%
\lambda\left\langle \left\langle \mathrm{p}\uparrow\right|  \right|
\mathrm{tr}\left(  \partial_{0}A^{\dag}A\tau_{3}\right)  \left|  \left|
\mathrm{p}\uparrow\right\rangle \right\rangle \nonumber\\
& =\frac{0.09}{2\pi}\frac{e}{F_{\pi}}\left\langle \left\langle \mathrm{p}%
\uparrow\right|  \right|  J_{3}\left|  \left|  \mathrm{p}\uparrow\right\rangle
\right\rangle \\
& =3.0\times10^{-4}\;,\nonumber
\end{align}
which is identical to that of ANW. On the other hand $\left(  \mu
_{I=1}\right)  _{3}$ is the same as that of the meron case (\ref{mu of I=1
(meron)}).

We summarize all the results in table 1.%

\begin{table}[htbp] \centering
\renewcommand{\arraystretch}{1.5}%
\begin{tabular}
[c]{|l|l|l|}\hline
& meron & instanton\\\hline
isospin $I_{i}$ & $I_{i}^{\mathrm{ANW}}$ & $I_{i}^{\mathrm{ANW}}$\\\hline
spin $J_{i}$ & $J_{i}^{\mathrm{ANW}}+S_{i}$ & $J_{i}^{\mathrm{ANW}}+S_{i}%
$\\\hline
Hamiltonian $H$ & $\frac{1}{4\lambda}\left(  I^{2}+J^{2}-\frac{1}{2}%
S^{2}\right)  $ & $\frac{1}{4\lambda}\left(  I^{2}+J^{2}\right)  $\\\hline
$F_{\pi}/F_{\pi}^{\mathrm{ANW}}$ & 1.05 & 1\\\hline
$e/e^{\mathrm{ANW}}$ & 0.99 & 1\\\hline
$\mu_{I=0}/\mu_{I=0}^{\mathrm{ANW}}$ & 0.32 & 1\\\hline
$\mu_{I=1}/\mu_{I=1}^{\mathrm{ANW}}$ & $-\frac{1}{3}$ & $-\frac{1}{3}$\\\hline
\end{tabular}%
\caption{Comparison of the meron and the instanton cases with ANW. \label
{key}}%
\end{table}%

\section{Summary and discussions}

SU(2) Skyrme model reduces in the semiclassical approximation to the $S^{3}$
system. We have applied to this system the idea of inequivalent quantizations
of quantum mechanics on $S^{3}.$ Inequivalent quantizations are specified by
the ``induced spin''; $S=0,\frac{1}{2},1,\cdots$ and by the induced gauge
fields; trivial $A_{\mu}=0$, meron (ME), zero-size instanton (INS), etc. The
baryonic states N$=(\frac{1}{2},\frac{1}{2})$ and $\Delta=(\frac{3}{2}%
,\frac{3}{2})$ are present only in S=0 and $S=1$ cases. The $S=0$ case
corresponds to ANW analysis. The $S=1$ case has been discussed in Sec 4. We
compare the cases $S=0$(ANW), $S=1$(ME) and $\ S=1$(INS) in the following. In
$S=1$ case the ``induced spin'' contributes to the spin operator of the
skyrmion. The Hamiltonian, on the other hand, has the same expression for
$S=0$ and $S=1$(INS), and there appears S dependence for $S=1$(ME). Using the
experimental values for $M_{\mathrm{N}}$ and $M_{\Delta}$ we obtain the same
values for $F_{\pi}$, $e$ and the isoscalar part of the nucleon magnetic
moment in the cases $S=0$ and $S=1$ (INS), while in the case $S=1$(ME) these
values get modified. The isovector part of the nucleon magnetic moment, on the
other hand, gets modified in the cases $S=1$(ME) and $S=1$(INS), and has the
same values different from the case of $S=0$ (ANW).

Thus we see that for SU(2) Skyrme model to describe the baryonic states, three
cases $S=0$(ANW), $S=1$(ME) and $S=1$(INS) of quantization are possible. When
compared with experiment, $S=0$(ANW) case realizes the reality best. However,
from the viewpoint of the quantum mechanics on $S^{3}$, these three cases are
on the same footing and there is no way of determining which quantization has
to be adopted.

Since the Skyrme model stems from the nonlinear sigma model, it would be
interesting to trace the field theoretical origin of $L_{\mathrm{induced}}$.
An obvious candidate is the Wess-Zumino term which is also related to topology
of the field configuration space. The field theoretical origin of
$L_{\mathrm{induced}}$\ is important also for the argument related to $g_{A}$
, because we need to consider the axial vector current of the model. This
problem is left for future investigations.

\makeatletter
\renewcommand{\theequation}
{A.\arabic{equation}} \@addtoreset{equation}{section} \makeatother
\appendix

\section*{Appendix}

In this Appendix, we derive the commutation relations (\ref{commutation
ralations}) , using the Dirac's quantization method for the system (\ref{meron
Lagrangian}).

Rewriting the Lagrangian (\ref{meron Lagrangian}) as%

\begin{equation}
L=-M+2\lambda\left(  \dot{a}_{\mu}\right)  ^{2}+\mathcal{H}_{\mu}\dot{h}_{\mu
}+\mathcal{B}_{\mu}\dot{a}_{\mu}\;,
\end{equation}
where $\mathcal{H}_{\mu}=(2h_{3},2h_{2},-2h_{1},-2h_{0})$ and $\mathcal{B}%
_{\mu}$ has appeared in (\ref{definition of B}) , we treat $a_{\mu}$ and
$h_{\mu}$ as the dynamical variables with the constraints
\begin{subequations}
\begin{align}
\chi_{1}  & =a_{\mu}^{2}-1\approx0\;,\\
\chi_{3}  & =h_{\mu}^{2}-1\approx0\;.
\end{align}
Conjugate momenta for $a_{\mu}$ and $h_{\mu}$ are defined by
\end{subequations}
\begin{subequations}
\begin{align}
\Pi_{\mu}  & =\frac{\partial L}{\partial\dot{a}_{\mu}}=4\lambda\dot{a}_{\mu
}+\mathcal{B}_{\mu}\;,\\
\Pi_{\mu}^{h}  & =\frac{\partial L}{\partial\dot{h}_{\mu}}=\mathcal{H}_{\mu
}\;,\label{momentum of h}%
\end{align}
respectively. (\ref{momentum of h}) gives the primary constraints
\end{subequations}
\begin{equation}
\chi_{4+\mu}=\Pi_{\mu}^{h}-\mathcal{H}_{\mu}\approx0\;.
\end{equation}
Hamiltonian is given by
\begin{equation}
H_{0}=\pi_{\mu}\dot{a}_{\mu}+\pi_{\mu}^{h}\dot{h}_{\mu}-L=M+\frac{1}{8\lambda
}\left(  \pi_{\mu}-\mathcal{B}_{\mu}\right)  ^{2}\;.
\end{equation}
Then consistency condition
\begin{equation}
\dot{\chi}_{1}=\left\{  \chi_{1},H\right\}  \approx0
\end{equation}
leads to the secondary constraint
\begin{equation}
\chi_{2}=a_{\mu}\pi_{\mu}\approx0\;.
\end{equation}
No further constraints follow from other consistency conditions.

We define the matrix $C_{ij}$ ($i,j=1,\cdots,7$) by $C_{ij}\equiv\left\{
\chi_{i},\chi_{j}\right\}  $ in terms of Poisson bracket $\left\{
\;,\;\right\}  $. Then, $\det C_{ij}=0$ shows that there exists first class
constraint among the above seven constraints $\chi_{i}$ ($i=1,\cdots,7$).

In order to construct the Dirac bracket, we introduce one gauge fixing
condition
\begin{equation}
\chi_{8}=h_{0}-1\approx0\;.
\end{equation}
Then, all the constraints $\chi_{\alpha}$ ($\alpha=1,\cdots,8$) become second
class; $\det$ $C_{\alpha\beta}=\det\{\chi_{\alpha},\chi_{\beta}\}\neq0$
($\alpha,\beta=1,\cdots,8$),
\begin{equation}
C_{\alpha\beta}\equiv\left\{  \chi_{\alpha},\chi_{\beta}\right\}  =\left(
\begin{array}
[c]{cccccccc}%
0 & 2 & 0 & 0 & 0 & 0 & 0 & 0\\
-2 & 0 & 0 & 0 & 0 & 0 & 0 & 0\\
0 & 0 & 0 & 2h_{0} & 2h_{1} & 2h_{2} & 2h_{3} & 0\\
0 & 0 & -2h_{0} & 0 & 0 & 0 & -4k & -1\\
0 & 0 & -2h_{1} & 0 & 0 & -4k & 0 & 0\\
0 & 0 & -2h_{2} & 0 & 4k & 0 & 0 & 0\\
0 & 0 & -2h_{3} & 4k & 0 & 0 & 0 & 0\\
0 & 0 & 0 & 1 & 0 & 0 & 0 & 0
\end{array}
\right)  \;.
\end{equation}
Inverse of $C_{\alpha\beta}$ is
\begin{equation}
C_{\alpha\beta}^{-1}=\left(
\begin{array}
[c]{cccccccc}%
0 & -\frac{1}{2} & 0 & 0 & 0 & 0 & 0 & 0\\
\frac{1}{2} & 0 & 0 & 0 & 0 & 0 & 0 & 0\\
0 & 0 & 0 & 0 & 0 & 0 & -\frac{1}{2h_{3}} & \frac{2}{h_{3}}k\\
0 & 0 & 0 & 0 & 0 & 0 & 0 & 1\\
0 & 0 & 0 & 0 & 0 & \frac{1}{4k} & -\frac{1}{4}\frac{h_{2}}{kh_{3}} &
\frac{h_{2}}{h_{3}}\\
0 & 0 & 0 & 0 & -\frac{1}{4k} & 0 & \frac{1}{4}\frac{h_{1}}{kh_{3}} &
-\frac{h_{1}}{h_{3}}\\
0 & 0 & \frac{1}{2h_{3}} & 0 & \frac{1}{4}\frac{h_{2}}{kh_{3}} & -\frac{1}%
{4}\frac{h_{1}}{kh_{3}} & 0 & -\frac{h_{0}}{h_{3}}\\
0 & 0 & -\frac{2}{h_{3}}k & -1 & -\frac{h_{2}}{h_{3}} & \frac{h_{1}}{h_{3}} &
\frac{h_{0}}{h_{3}} & 0
\end{array}
\right)  \;.
\end{equation}
The definition of Dirac bracket
\begin{equation}
\{A,B\}_{\mathrm{D}}\equiv\{A,B\}-\{A,\chi_{\alpha}\}C_{\alpha\beta}%
^{-1}\{\chi_{\beta},B\}
\end{equation}
\smallskip gives the relations
\begin{align}
\left\{  a_{\mu},a_{\nu}\right\}  _{\mathrm{D}}  & =0\;,\label{Dirac brackets}%
\\
\left\{  a_{\mu},\pi_{\nu}\right\}  _{\mathrm{D}}  & =\delta_{\mu\nu}-a_{\mu
}a_{\nu}\;,\nonumber\\
\left\{  \pi_{\mu},\pi_{\nu}\right\}  _{\mathrm{D}}  & =-a_{\mu}\pi_{\nu
}+a_{\nu}\pi_{\mu}\;,\nonumber\\
\left\{  S_{i},S_{j}\right\}  _{\mathrm{D}}  & =\varepsilon_{ijk}%
S_{k}\;,\nonumber\\
\left\{  a_{\mu},S_{i}\right\}  _{\mathrm{D}}  & =\left\{  \pi_{\mu}%
,S_{i}\right\}  _{\mathrm{D}}=0\;,\nonumber
\end{align}
where $S_{i}\equiv-\mathrm{tr}\left(  T_{i}hKh^{-1}\right)  $ is the ``induced
spin'' variable. From (\ref{Dirac brackets}), commutation relations for the
system (\ref{meron Lagrangian}) become (\ref{commutation ralations}).

\bigskip

{\Large Acknowledgments} We would like to thank K.Hasebe for discussions and
comments.

\end{document}